# Impact of Mobile Transmitter Sources on Radio Frequency Wireless Energy Harvesting

Antonio Hernandez Coarasa[1], Prusayon Nintanavongsa[1], Sugata Sanyal[2], and Kaushik R. Chowdhury[1]
[1] Department of Electrical and Computer Engineering, Northeastern University, Boston, MA, USA
[2]Corporate Technology Office, Tata Consultancy Services, India

***Abstract*—Wireless energy harvesting sensor networks constitute a new paradigm, where the motes deployed in the field are no longer constrained by the limited battery resource, but are able to re-charge themselves through directed electromagnetic energy transfer. The energy sources, which we call actors, are mobile and move along pre-decided patterns while radiating an appropriate level of energy, sufficient enough to charge the sensors at an acceptable rate. This is the first work that investigates the impact of energy transfer, especially concerning the energy gain in the sensors, the energy spent by the actors, and the overall lifetime in the resulting mobile sensor-actor networks. We propose two event-specific mobility models, where the events occur at the centers of a Voronoi tessellation, and the actors move along either (i) the edges of the Voronoi cells, or (ii) directly from one event center to another. We undertake a comprehensive simulation based study using traces obtained from our experimental energy harvesting circuits powering Mica2 motes. Our results reveal several non-intuitive outcomes, and provide guidelines on which mobility model may be adopted based on the distribution of the events and actors.**

## I. Introduction

Powering battery constrained sensors with energy harvesting (EH) has resulted in a new paradigm of long-lived wireless sensor networks (WSNs). Such sensors may rely on external and possibly ambient sources of energy, such as the sun, wind, naturally occurring vibrations, among others, and convert these forms of energy into useful electrical energy that is stored in a capacitor for later use. However, these sources exhibit spatial and temporal variations in the energy that is actually incident on the harvesting circuits, which makes complete dependence on these sources a major concern. Recently, we demonstrated a new technique of powering sensors through electromagnetic radiation in the radio frequency (RF) range [1], which can result in a directed energy transfer. The aim of this paper is to investigate scenarios where the source of energy is mobile, and has power control. Thus, how to move along Event Points (EP) in a WSN while ensuring maximum energy transferred to the sensors in need is the topic of focus in this work.

In the rest of this paper we use the term *actor* to indicate an energy-rich source, which is mobile and can move around in the network. It radiates energy through RF transmissions, which is captured and converted by the on-field sensors connected to energy harvesting circuits.

In the architecture considered in this paper, the *actors* move under different mobility models. They also radiate power at different levels depending upon the distance from the event. We assume that the sensors around the event location are maximally impacted by the event, i.e., they perform tasks of sensing, reporting the readings, compressing measurements based on correlation and aggregating the data from neighbor. These activities not only involve higher transmission costs, but also higher expenditure from on-board computations. Thus, the primary aim of the *actor* is to ensure that the nodes around the event are kept alive, and any variation in the radiated power is always bounded by the minimum RF power level incident at these event locations. Moreover, as the *actors* move, they themselves consume energy, and path planning needs to be carefully considered in the design. In this study, we look at an in-depth evaluation of multiple additional factors including the effect of sensor duty cycles, amount of *actors*, number of event locations, the minimum required power to charge for a given sensor, the density of sensor deployment, and the frequency in which the radiation occurs.

In summary, the main contributions of this study are:

- We explore the tradeoff between (i) whether to transmit at high power from a distance, or (ii) move closer to the event area to decrease the required power to transmit, with the resulting impact on the energy loss due to motion.
- We earlier designed and interfaced two prototypes that harvest energy from licensed in the 642 MHz, and the easily accessible ISM bands [1]. Here, we study if the energy transfer efficiency in the licensed frequencies justifies the additional licensed user avoidance overhead.
- We identify which of the environmental factors (e.g., node density, event density, actor density, mobility pattern, transmission power variation) are dominant in ensuring long network lifetime.

The remainder of this paper is organized as follows. In Section II we describe the related work. In Section III we describe the network setup and the two mobility models. The main body of the paper is the extensive performance evaluation study in Section IV. Finally, we draw conclusions in Section V.

## II. Related Work

Energy harvesting from RF waves constitutes a new paradigm [2] that goes beyond the commonly assumed forms of energy obtained from wind [3] and the incident sunlight [4]. The viability of this technology has been demonstrated through different commercial and research prototypes [5, 6, 7, 8], apart from our own efforts in [1]. The overall aim remains to obtain enough energy to charge a capacitor up to $1-3$ V that can run a low-power sensor mote. The concept of *actors* that react to events and address them has been explored in [9]. Our actors are mobile and enabled with a perennial source of power. These actors may move under a variety of mobility patterns. For the purpose of this paper, we assume that the actors move along certain specific paths, based on where the events actually occur. This allows focused charging of the sensors at those event locations. We make the use of Voronoi tessellations in this work, where the area is split into regions, called as Voronoi cells [10].

## III. Energy transfering through Mobile Actors

Our proposed method of energy transferring relies on *actors* that move along a region that is partitioned into Voronoi cells. First, we describe how these cells and paths are constructed.

Let $S = \{p_1, p_2, ..., p_i, ..., p_n\}$ the set of the points that correspond to specific event locations in the region of interest. These event locations typically signal a feature of interest, such as an extreme temperature, that requires continuous sensing and transmission for the sensors close to that point. Let $V(p_i)$ denote the set of all sensors that are closer to the event point $p_i$, than any other point belonging to $S$.

$$V(p_i) = \{x : |p_i - x| < |p_j - x|, \forall j\; j= i\} \qquad (1)$$

Next, we describe the two mobility models that we shall use in this investigation.

### A. Mobility Models for the Actors

In the first mobility model, called *center-to-center mobility model* (CM), the actors move along paths that connect the EPs. We use the traveling salesman problem algorithm to construct the connected paths from one event point to the next, so all the points are eventually traversed (see Figure 1a). Here, the focus is to ensure that the sensors close to the Voronoi cell centers, i.e., the respective EPs of the Voronoi cells, have the maximum possible lifetime. However, this "event-centric" energy transfer may not be representative of a wider class of WSN applications, where multiple nodes forward data packets towards a sink. Thus, not only nodes close to the EP, but also in the peripheral region need to be actively charged.

In the second mobility model, called as *around edges moving model* (EM), the actors move along the edges of the Voronoi cells. The key aspect of using the edge is that the energy transfer occurs on a much wider extent, covering those nodes that may potentially be farther away from the event point. However, such nodes may well participate in data forwarding, and need to be charged as long as the sensors close to the event generate readings. Assuming that actors are initially assigned to individual vertices of the Voronoi cells, they continue to move back and forth along all edges that intersect that vertex. In Figure 2a, the motion is showed by the bold dark line, for the rightmost vertex of the tessellation.

### B. Initial Deployment of the Actors

While the mobility model is mainly responsible for assured transfer of energy, the initial deployment has significant bearing on the efficiency of this transfer. If there is a mass concentration of actors in any one location, then the other areas of the network get starved of future energy transfers. Hence, we delve into the deployment issue is detail.

First, we present a case for the deployment of actors for the CM model. The steps are as follows:

- *Step1:* If the number of actors is greater than the number of EPs, first, one actor is deployed in each EP in turn
- *Step2:* Repeat *Step1* until the remaining actors are less than the EPs (see Figure 1a that shows an example of 4 EPs and 11 actors)
- *Step3:* Once all the actors are deployed in the EPs they are re-positioned equidistantly from each other on the edge connecting two successive EPs belonging to the traversal path (see Figure 1b)

Thus, we get a final distribution of actors with the maximum coverage along the traversal edges.

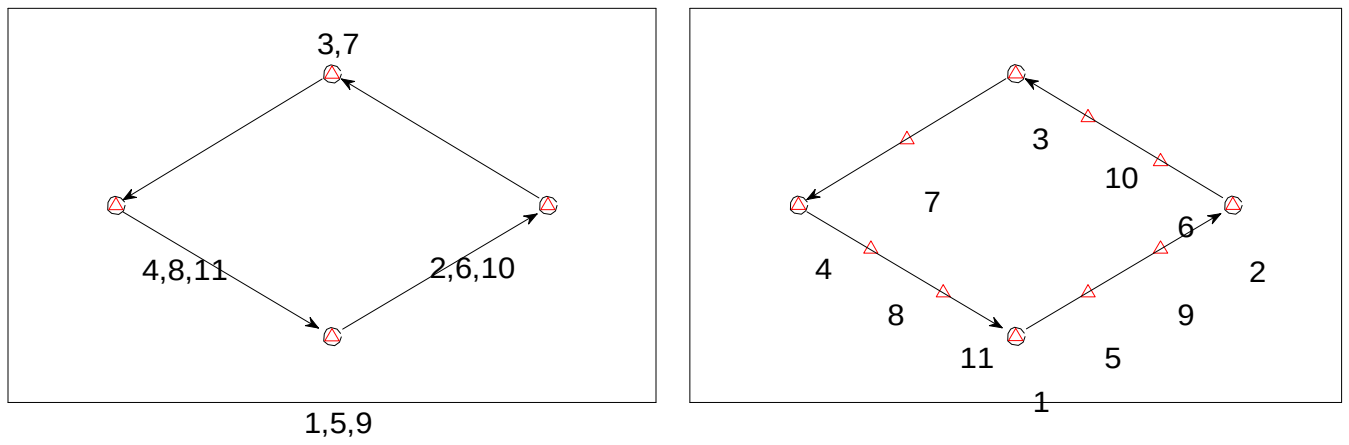

(a) Starting distribution among EP (b) Final equidistant distribution

Fig. 1: Actor deployment for CM

For the actor deployment in the EM model, we outline the following steps.

- *Step1:* If the number of actors is greater than the number of inner Voronoi corners, first, one actor is deployed in each corner
- *Step2:* Repeat *Step1* until the remaining actors are less than the inner corners (see Figure 2a that shows an example of 4 EPs and 11 actors)
- *Step3:* Once all the actors are deployed in the corners they are redistributed in the different branches following the same method, such as each branch has the same number of actors (or close)
- *Step4:* Finally, each actor belonging to a given branch is re-positioned equidistantly from each other connecting the two successive inner corners (see Figure 2b)

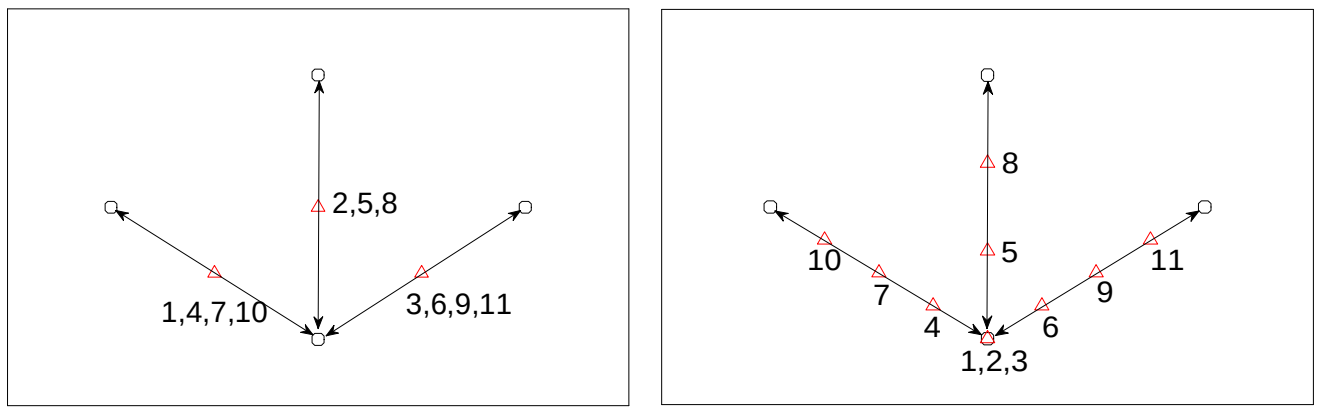

(a) Starting distribution among edges (b) Final equidistant distribution

Fig. 2: Actor deployment for EM

### C. Moving actors and energy propagation

We let the actors move with constant speed along the chosen paths. Irrespective of the mobility model used, the actors radiate just enough energy to charge the sensors close to the events, i.e., around the EPs, considering a minimum required power at the EP. Also, the energy consumed by an actor for its physical movement at a speed of 2 $m/s$ is 150 $mW$, given by [9]:

$$E_V(W) = 0.05\, W/(\text{m/s})^{\gamma}\ ,\ \gamma = 1.5 \qquad (2)$$

To compute the required energy to be transmitted by an actor to provide the energy level requested around the EP, the Friis equation (3) is used. R is the emitter-receiver distance. This equation is a function of the frequency (here, expressed through the wavelength $\lambda$) as it varies from $f$ = 642 MHz ($\lambda$ = 46.7 cm) to $f$ = 915 MHz ($\lambda$ = 32.8 cm). Consequently, the required energy increases proportional to the square of the frequency variation. Due to the limitation of the harvesting circuits, any power received under −20 dBm has been null. Friis equation, altogether with unitary antenna gains and isotropic antenna propagation are used here to simplify the simulation. Also multipath effect is neglected for tractability of the simulation.

$$P_r(W) = P_t(W) \times G_t \times G_r \times \left(\frac{\lambda}{4\pi R}\right)^2\ ,\ \lambda = \frac{c}{f} \qquad (3)$$

## IV. Performance Evaluation and Observations

### A. Simulation Setup

We use MATLAB to study the impact of mobility in this work, with all the parameters chosen for MICA2 motes and our harvesting platform characteristics [1].

We use the stop condition for the simulation as follows: When the sensor coverage reduces to 50% of the area of deployment, with 5 m sensing radius of each node (i.e., as nodes start dying owing to energy loss), the simulation for that run is stopped. All calculations of residual energy mentioned in this section are obtained from the average values of the sensors closest to their respective EP. The power consumed by actors is calculated as the average over the cycles (time slots) of the sum of all the actors' individual consumption.

Table I shows the default parameters used in this paper.

| Parameter | Default | Range |
|---|---|---|
| Probability of consuming | 1/30 | 1/[10,20,30,40] |
| Max TX power | 36 | 36 dBm |
| No. of actors | 10 | 10,20,30,40 |
| No. of Event Points | 10 | 10,20,30,40 |
| Min required power at EP | -5 | -20,-10,0,5,10 dBm |
| Harvesting Frequency | 915 | 642MHz,915MHz,2.4GHz,5.1GHz |
| Area | 200 | 150,200,250 $m^2$ |

TABLE I: Parameters for the simulations

### B. Observations

As we evaluate each parameter, we vary it for four different types of mobility models (shown on the X-axis), i.e., (i) static actors placed along the edges in the EM case (called as *static-EM*), (ii) mobile actors for the EM case (called as *mobile-EM*), (iii) static actors placed on the EPs, for the CM case (called as *static-CM*), and (iv) mobile actors moving from one EP to another, again for the CM case (called as *mobile-CM*). We evaluate the impact of mobility with respect to the following parameters:

*1) Minimum received power at EP:* Figure 3a shows the energy consumption of actors with various values of the minimum required power at the EP. It is clear that both the cases of static and mobile EM show little deviation with the change in this power requirements at the EP. In the EM case, the actors are rarely close to the EP, and thus the actors are always forced to transmit at a higher power than the minimum required level. On the other hand, for the CM case, actors pass really close (over the point), and hence, at some instances, the considered transmission power can be drastically reduced, producing a bigger variation.

Figure 3b shows the residual energy of sensors. Overall, mobile scenarios indicate an increase of residual energy while it is almost constant in static models. Also, there is considerable increment in residual energy in mobile CM compared to mobile EM. This is because, in mobile CM, sensors get a higher recharging rate when the actors are close to the EP and transmitting at a high power.

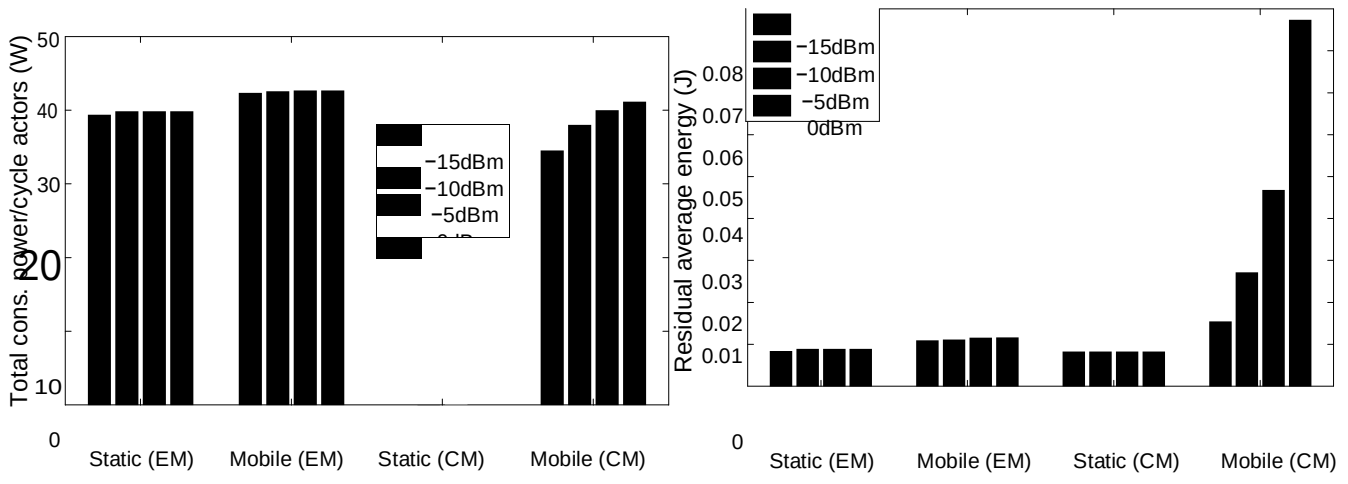

(a) Total actors consumed energy/cycle (b) Average sensors residual energy

Fig. 3: Results under required power level variation

*2) Frequency of energy transmission for the actors:* The effect of transmission frequency on the energy consumption of actors and the residual energy of sensors are shown in Figures 4a and 4b, respectively. In Figure 4a, it is clear that the influence on the increment of energy consumed by the actors is not substantial (3% increment from 642 MHz to 915 MHz). On the contrary, the improvement over the residual energy level on sensors is drastic (50% increment from 915 MHz to 642 MHz mobile EM). This deviation is larger in CM as the actors pass very close to the sensors used for residual energy calculations. On the other hand, in EM, all the sensors located around the middle regions of the Voronoi cell get a good average charging rate, as shown in Figure 4b.

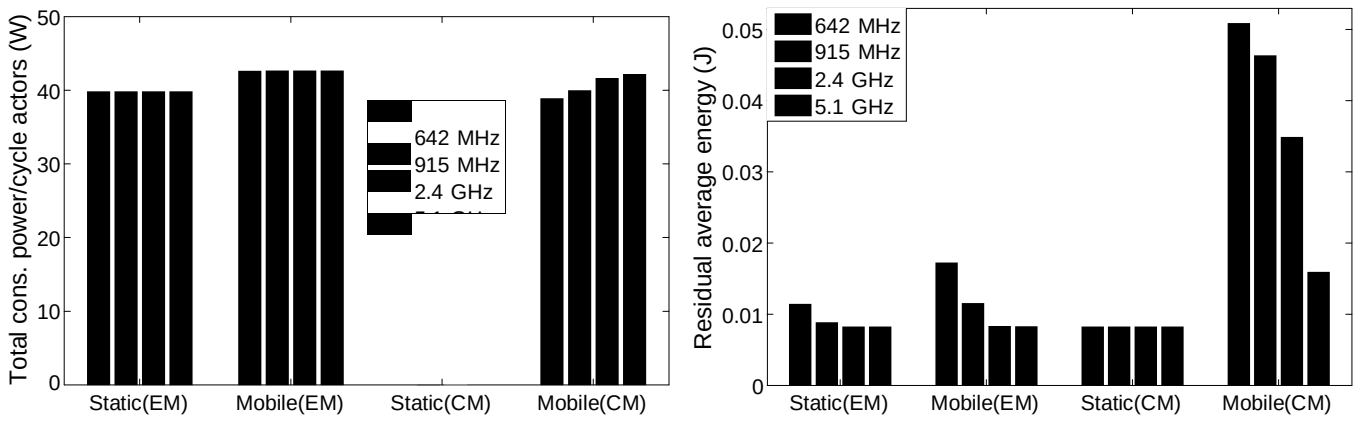

(a) Total actors consumed energy/cycle (b) Average sensors residual energy

Fig. 4: Impact of frequency variation

Figures 5a and 5b show the energy consumption of actors and the residual energy of sensors of mobile EM, respectively, with both transmission frequency and minimum required power at the EP.

With low minimum required power requested (at −20 dBm), there exists a large variation in energy consumption of actors while the residual energy of sensors tends to be somewhat constant. The constant level of the residual energy is because all the received energy is more influenced by actors' power level than the path loss. The large variation in consumed energy is due to different level of effort undertaken by the actors to allow the sensors around the EP to receive the required power. The difference in consumed power is approximately 3 times, from 642 MHz to 5.1 GHz.

With higher minimum required power (at 20 dBm), the residual energy of sensors shows a large variation, while the consumed energy asymptotically converges for the actors. The large variation in residual energy is a result of the pathloss effect. The consistence in consumed energy is because actors are forced to transmit with the highest power allowed, omitting the required power by the EP.

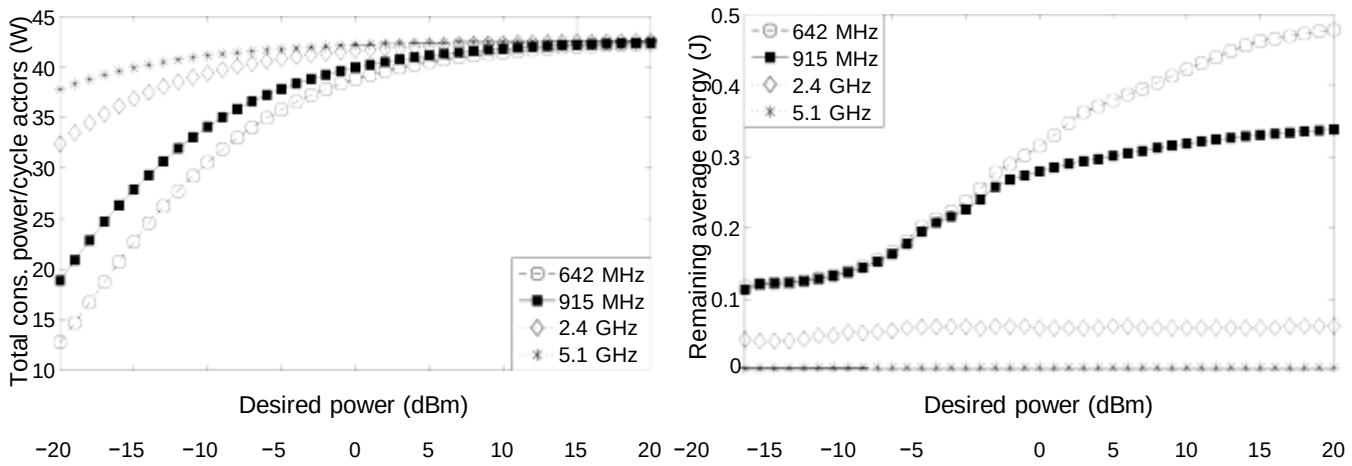

(a) (b)

Fig. 5: Consumed energy (a), and Residual energy (b) measured in Joules (Y axis) for different frequencies when the minimum required power at the EP (X axis) varies

*3) Number of actors:* Figures 6a and 6b show the consumed energy of actors and residual energy of the sensors.

At first glance, the increase in the number of actors seems to give a linear increment on the residual energy, and there exists a strong dependency within a moving model. In the CM case, as the actors move in the same direction at all time, this expected result is intuitive. However, in EM, actors go back and forth from their original position to the neighboring Voronoi corners. Consequently, increasing number of actors does not guarantee an increase in both consumed and residual energy. It is also clear that the increase in energy consumed in CM is larger, while the variation in EM tends to be smaller.

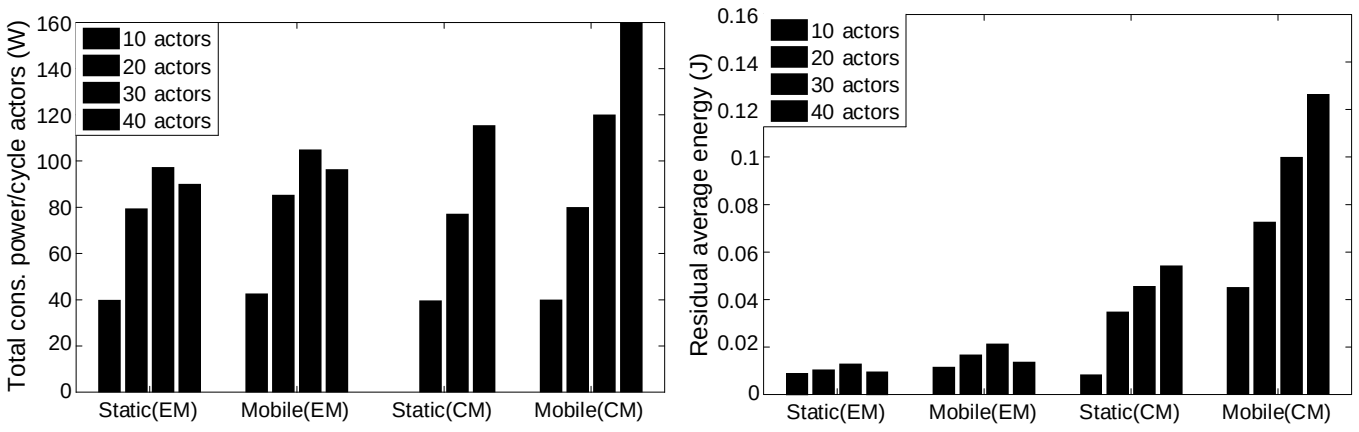

(a) Total actors consumed energy/cycle (b) Average sensors residual energy

Fig. 6: Impact of varying number of actors

*4) Number of event points:* As we vary the number of event points, i.e., the EPs, the number of corresponding Voronoi cells also changes. Thus, as EPs increase, on one hand the overall length of edges that the actors need to travel increases, while on the other hand the inter-event distance decreases. Thus, the actors spend reduced amount of transmission energy for re-charging sensors around the EP (though mobility-caused energy consumption by the actors is higher). Additionally, as the actors move along the edges back and forth that intersect the initial deployment vertex, on average, higher amount of actors are needed for the network. These observations are evident in Figure 7a where we see how the energy consumption is impacted, when we vary the number of EPs is

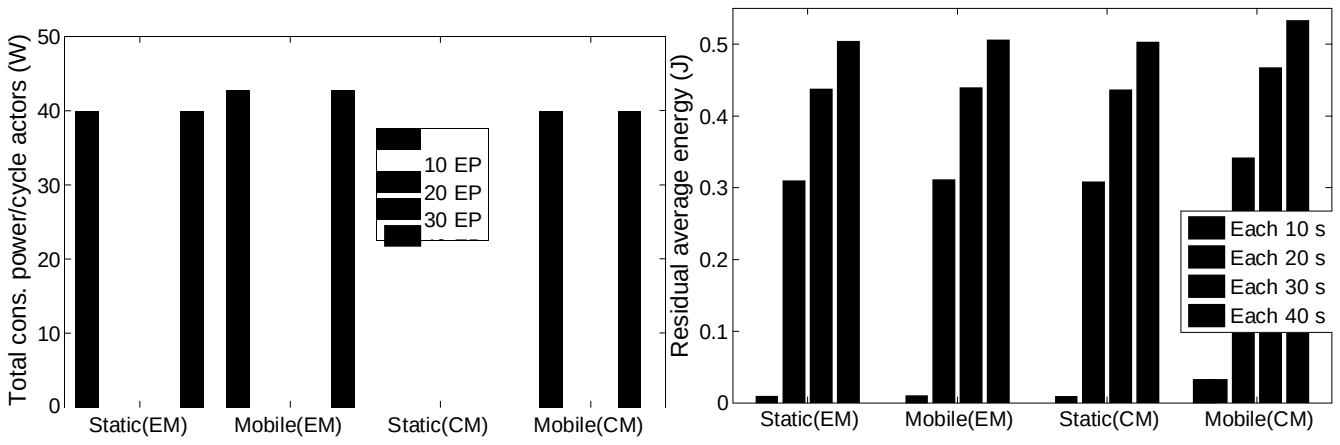

(a) Consumed energy/cycle for all actors (b) Average sensor residual energy

Fig. 7: Consumed energy when varying EPs with 10 actors (a). Residual energy level when varying probability of energy consumption by sensors (b)

*5) Probability of sensor energy consumption:* This parameter, whose purpose is to evaluate how the sensor sleep-awake cycle influences the network, is the inverse of average Sleeping time of the sensors (displayed in Table I). The energy spent by the actors is kept constant, independent of the probability of energy consumed by sensors. Figure 7b indicates how the residual energy of sensors increases significantly, and the significant variation suggests it is a good parameter to be tuned in order to increase the network lifetime

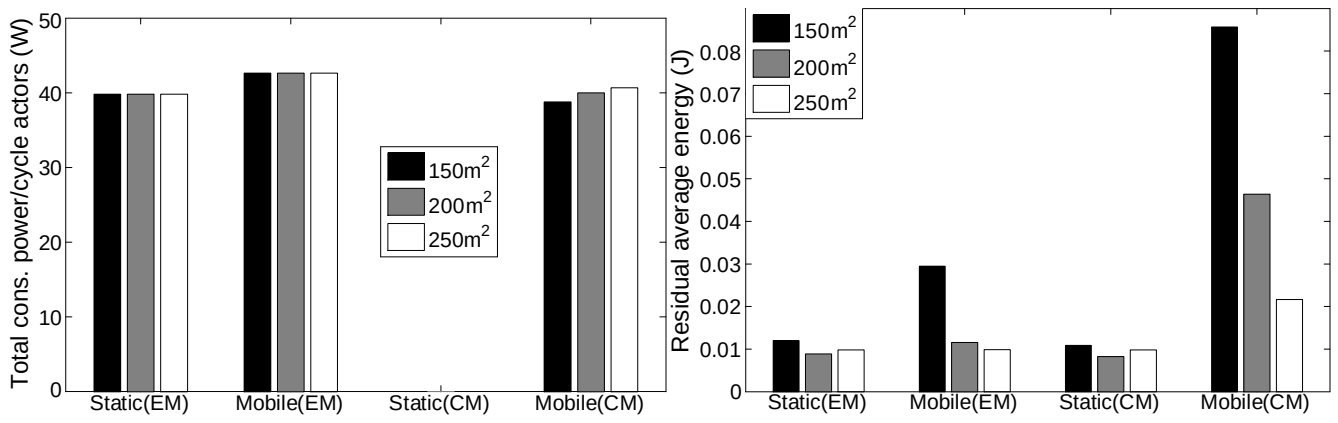

(a) Total actors consumed energy/cycle (b) Average sensors residual energy

Fig. 8: Impact of varying the deployment area

*6) Area scenario:* We initially performed the simulation varying the measurement area with a fixed number of sensors. This resulted in an abrupt energy depletion of some sensors due to sparse sensor density, and thus, terminating the simulation prematurely. Thus, we determined it was crucial to maintain a constant sensor density so that the simulation delivers comparable results.

Figures 8a and 8b show the consumed energy of actors and residual energy of sensors, respectively, with various area sizes. While the increment of the consumed energy of actors negligible, the residual energy of sensors tends to increase up to 4 times as the deployment area is varied from 250 $m^2$ to 150 $m^2$ for mobile CM, and up to 3 times for mobile EM for the same conditions.

## C. Network lifetime

We next investigate the improvement in network lifetime (all results are in the unit of seconds) with respect to various parameters as mentioned below:

- Frequency variation: Figure 9a shows the effect of various transmission frequencies on the network lifetime. Following the pathloss equation, the lower frequencies are less susceptible to signal attenuation. This directly results in stronger received signal, and hence, prolonged network lifetime.

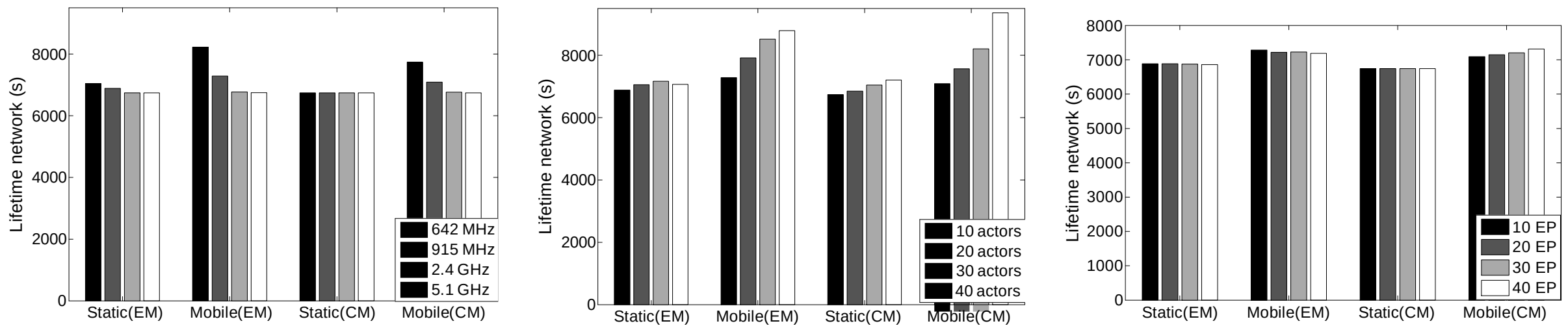

(a) Energy transferring frequency (b) Number of actors (c) Number of event points (EPs)

Fig. 9: Impact of transmission frequency, number of actors and event points on network lifetime

- Increasing of number of actors: Figure 9b shows the effect of number of actors on network lifetime. It is clear that increasing number of actors increases the network lifetime. Moreover, the mobile movement model delivers a smoother improvement than the static movement model.
- Number of event points: Figure 9c shows the effect of number of event points on network lifetime. Non intuitively, increasing number of event points yields an improvement in terms of energy consumption, but there is a negligible improvement in network lifetime with this increase. The reason is as follows: The energy transmitted by the mobile nodes is utilized more efficiently with more sensors requiring energy (the total time for energy transmission remains same in all cases). However, the extra load of charging these multiple event points increases the traversal time of the mobile actors, and overall, there is no improvement in the lifetime.

## V. Discussions and Conclusions

From our investigation of the various mobility models, there are certain scenarios that are particularly suited for either the CM or the EM, which we summarize below:

- The sensors have a certain minimum incident power requirement for charging the storage capacitor. This threshold power level does not significantly impact the EM case. However, judicious selection of this level becomes very important for the CM case, which impacts the storage energy drastically (e.g., In CM for a variation of 5 dBm in the minimum required power for re-charging, the percentage increase in the storage energy is over 85% for the entire network).
- The lower frequency bands have a significant improvement in the network lifetime. Importantly, there is a large difference (e.g., 50% when using the EM case) in the residual energy levels for the sensors with the use of lower frequencies (when using channels in the 600 MHz band, over 900 MHz).
- The increase in the number of energy transferring actors has a direct influence in CM case, while for EM; it depends predominantly on the geometry of the scenario. This is because the path traversal length for typical deployment scenarios often outweighs the gain from multiple active actors.
- The duration of the awake time for the sensors impacts the network lifetime to a large extent in the case of CM. For e.g., a variation from 20 to 30 seconds gives an increase of 35% in the residual energy level, and 30% improvement in the lifetime, while maintaining a constant energy consumption ratio.

We conclude with the general assessment that the CM gives better performance in small deployment scenarios, and when there is a higher density of sensors concentrated around the event points. The EM provides better results on large deployment scenarios where the sensor density is lower and the events are scattered, with higher separation distance between them.

## Acknowledgment

This material is based upon work supported by the US National Science Foundation under Grant No. CNS-1143681.